# COVID-MTL: Multitask Learning with Shift3D and Random-weighted Loss for Automated Diagnosis and Severity Assessment of COVID-19

Guoqing Bao[1, *], Huai Chen[2], Tongliang Liu[1], Guanzhong Gong[3], Yong Yin[3], Lisheng Wang[2] and Xiuying Wang[1, *]

[1]School of Computer Science, The University of Sydney, J12/1 Cleveland St, Darlington, Sydney, NSW, 2008, Australia
[2]Department of Automation, Institute of Image Processing and Pattern Recognition, Shanghai Jiao Tong University, Shanghai, China
[3]Department of Radiation Oncology, Shandong Cancer Hospital and Institute, Shandong First Medical University and Shandong Academy of Medical Sciences, Jinan 250117, China.
*Correspondence: guoqing.bao@sydney.edu.au; or xiu.wang@sydney.edu.au

*Abstract* – There is an urgent need for automated methods to assist accurate and effective assessment of COVID-19. Radiology and nucleic acid test (NAT) are complementary COVID-19 diagnosis methods. In this paper, we present an end-to-end multitask learning (MTL) framework (COVID-MTL) that is capable of automated and simultaneous detection (against both radiology and NAT) and severity assessment of COVID-19. COVID-MTL learns different COVID-19 tasks in parallel through our novel random-weighted loss function, which assigns learning weights under Dirichlet distribution to prevent task dominance; our new 3D real-time augmentation algorithm (Shift3D) introduces space variances for 3D CNN components by shifting low-level feature representations of volumetric inputs in three dimensions; thereby, the MTL framework is able to accelerate convergence and improve joint learning performance compared to single-task models. By only using chest CT scans, COVID-MTL was trained on 930 CT scans and tested on separate 399 cases. COVID-MTL achieved AUCs of 0.939 and 0.846, and accuracies of 90.23% and 79.20% for detection of COVID-19 against radiology and NAT, respectively, which outperformed the state-of-the-art models. Meanwhile, COVID-MTL yielded AUC of 0.800 ± 0.020 and 0.813 ± 0.021 (with transfer learning) for classifying control/suspected, mild/regular, and severe/critically-ill cases. To decipher the recognition mechanism, we also identified high-throughput lung features that were significantly related ($P < 0.001$) to the positivity and severity of COVID-19.

## 1. Introduction

As firstly reported in December 2019 [1], COVID-19 was identified as a novel coronavirus (SARS-CoV-2) with severe respiratory symptoms similar to pneumonia and seasonal flu, such as fever, cough, fatigue, and myalgia [2]. The outbreak of the disease has triggered the World Health Organization (WHO) to declare it as a pandemic. This human-to-human transmission disease has resulted in more than 80 million infections worldwide with over 1.7 million deaths as of December 2020 according to statistics released by Johns Hopkins University.

To constraint the spread of COVID-19 pneumonia, besides personal protection, WHO recommended preventive measures including quickly identifying suspect cases, timely testing, isolating infectious people, and more importantly, identifying all close contacts of the infected [3]. Nucleic acid test (NAT) via real-time polymerase chain reaction (RT-PCR) is considered as an operational "gold standard" for detection of the causative agent of COVID-19 [4, 5]. However, the RT-PCR test suffers from a high false-negative rate especially in initial disease presentation and asymptomatic people [6, 7] which may due to prolonged nuclei acid conversion, lack of sufficient test kits, and the low quality of the swab samples [5, 6, 8].

It was reported that SARS-CoV-2 affects lung lobes and patients infected with COVID-19 pneumonia are widely exhibited ground-glass opacities (GGO), consolidation, or both in their chest computer tomography (CT) scans [8-10]. Naturally, such anatomical changes can also be captured by measuring imaging features, especially texture features, and used for COVID-19 diagnosis. Besides, the chest CT scan is suggested to be able to detect COVID-19 in the early stage, especially useful for screening asymptomatic patients or patients with negative NAT results [11, 12]. More importantly, signs of disease progression can also be observed from chest CT images, and as reported [13] that GGO, GGO plus reticular pattern or consolidation were all common in the early rapid progressive stage, GGO plus consolidation dominated the advanced stage, and GGO plus consolidation sharply decreased in the recovery (absorption) stage. Consequently, the chest CT scan has become a complementary strategy of NAT and is widely used in clinical practice.

Considering the widespread of COVID-19 across 191 countries/regions, the rapid increase of the number of new cases, and the success of deep learning in medical image analysis, there is an urgent need to develop a deep learning-based system for automated assessment of COVID-19. Several methods have been proposed and achieved promising results. For instance, Li et al. [14] proposed a ResNet50-based COVNet model, from which a series of CT image slices were feed into different network branches, and the feature maps obtained from individual branches were finally concatenated, for detection of COVID-19 and community-acquired pneumonia. Later, Harmon et al. [12] proposed an artificial intelligence (AI) system to detect COVID-19 pneumonia using multinational chest CT datasets, which achieved up to 90.8% accuracy. Similar performance was achieved by Sun et al. [15] and in this method, they first extracted imaging features from volumetric CT scans and proposed a deep forest network guided by adaptive feature selection for COVID-19 classification. Wang et al. [16] developed a tailored 2D convolutional neural network, named COVID-Net for the detection of COVID-19 using chest X-ray images. This model was later redesigned in [17] for COVID-19 CT image slice classification. More recently, Shorfuzzaman and Hossain [18] used a fine-tuned pre-trained convolutional encoder to capture feature representations of COVID-19 from limited X-ray training samples, and then adopted a Siamese network for classification of COVID-19. Besides the binary classification of COVID-19, Wang et al. [19] developed a deep learning model that can simultaneously localize the infectious regions of COVID-19 on chest X-ray images. Tang et al. [20] extracted radiomic features from CT images and then combined them with clinical indices for classification of severe vs. non-severe COVID-19 in a small cohort using a random forest model. Very recently, Ning et al. [21] made their COVID-19 dataset publicly available and proposed to use CT imaging data as well as clinical features for detection and severity assessment of COVID-19. They have manually labeled 19,685 CT slices to train their single-task CNN models.

The existing works mainly focused on detecting COVID-19 using CT or X-ray images against either radiological or NAT results. Given the high false-negative rate of NAT (somewhat between 2% and 33% in repeat sample testing [5]), the prediction results of some existing solutions might have biases. More seriously, though people with NAT-negative infection may not present any symptoms, they still carry the virus and may have a big chance to transmit among their contacts, which thereafter poses an even greater risk to the communities since it is harder to do contact tracing for asymptomatic transmissions. Radiology, especially computer tomography, which is more sensitive than other imaging modalities for early diagnosis of COVID-19, is now served as an essential complementary method. Thus, automated detection of COVID-19 against both NAT and computer tomography is increasingly needed and may be more useful in clinical practice. While, most of the existing CT solutions were slice-based, which may not be applicable in mass practice because the selection of proper CT slices for model inference still requires expert involvements. Besides diagnosis, automated and fast severity assessment of COVID-19 may be especially beneficial for severe patients given the extreme shortage of hospital beds to handle the unexpected surge of COVID-19 admissions across many countries.

To address these challenges, we propose a multitask-learning (MTL) framework to ensemble 3D CNN and auxiliary feed-forward neural network (FNN) to harness volumetric CT inputs and high-throughput CT lung features for automated and simultaneous detection of COVID-19 pneumonia against both radiology (diagnosed by radiologists using CT scans) and NAT (RT-PCR) as well as assessing the severity of the infection. Due to the imbalance of task difficulty, more difficult tasks such as severity assessment prone to slow-convergence in the MTL learning procedure. To tackle this issue, a novel random-weighted loss function is proposed to prioritize vulnerable COVID-19 tasks that aims to alleviate task dominance and enhance joint learning performance. COVID-MTL uses chest CT scans as inputs that are more stable compared to the slice-based approach since its inference process is fully automated. To overcome the hurdle faced by conventional 3D CNNs when processing volumetric CT inputs, we proposed a novel Shift3D that introduces space variances on low-level volumetric feature representations to alleviate overfitting and improve convergence and accuracy for state-of-the-art 3D CNN components. In addition, accurate lung segmentation is necessary for automated diagnosis, we thereby design a novel unsupervised algorithm to address the under-segmentation of crucial and diagnostic-relevant structures like GGO in COVID-19 CT scans.

## 2. Related work

### 2.1 Lung segmentation from COVID-19 CT scans

Lung segmentation is a necessary and critical step for the diagnosis and treatment of lung diseases, especially in the early stage. Conventionally, U-net, a symmetric model architecture that is widely used in medical image segmentation, is applied for lung and lung lesion/nodule segmentation [22, 23]. However, this method requires lung delineation masks that are paired to each input CT slice for training. Recent studies have shown that people infected with SARS-CoV-2 may undergo ground-glass opacities or GGO in their lungs within few weeks after symptom onset and thus subsequently demonstrate a white lung appearance in CT scans [9, 24]. The white lung areas may introduce additional difficulties for some of the existing lung segmentation methods, especially for algorithms that involving intensity or thresholding, where under-segmentation of GGO may occur.

To tackle those problems, different strategies have been proposed more recently. For instance, Oulefki et al. improved a multilevel thresholding algorithm based on Kapur entropy for automatic segmentation of COVID-19 infected lung regions from chest CT scans [25]; Fan et al. proposed a semi-supervised framework for segmentation of lung infections from COVID-19 CT scans, from which limited labeled images and randomly selected propagation strategies were used to train an Inf-Net CNN model [26].

However, given the shortage of radiologist engagement during the pandemic, an unsupervised segmentation algorithm is more favorable and more viable for mass studies and applications on COVID-19. In this research, we intended to improve a classical unsupervised lung segmentation algorithm [27], which was widely adopted by the community (e.g. Kaggle competition, Data Science Bowl 2017), for the following tasks, i.e. detection and severity assessment of COVID-19.

### 2.2 3D Convolutional Neural Network

The convolutional neural network was initially proposed to process 2D images, including handwriting recognition and natural image classification. Besides extracting features from the spatial dimensions, 3D convolution was later introduced to simultaneously handling the temporal dimensions of an input series, such as motion information captured from multiple adjacent video frames [28], hand pose signals that estimated from the single depth image [29], and organ tissue segmented from volumetric medical images [30]. Regarding processing CT images, one can either take a single CT slice as input using 2D CNNs which fail to leverage temporal context from adjacent slices, or leveraging inter-slice context from volumetric input by harnessing 3D convolution kernels. Although 3D CNNs can lead to improved performance in comparison to their 2D counterparts, the benefit comes with an extreme memory and computational cost due to the complexity of 3D convolution and increased number of network weights. To tackle the problem, well-known resource efficient 2D CNNs have been recently converted to 3D CNNs to leverage the capability of spatio-temporal features [31], such as ResNet3D, SqueezeNet3D, and MobileNet3D. SqueezeNet is one of the lightweight CNN architectures, which can achieve similar accuracy to AlexNet by only using 50 times fewer parameters [32]. In this work, we build our 3D CNN model with SqueezeNet as the backbone.

### 2.3 Multitask Learning

Multitask learning (MTL) is referred to as a learning paradigm that aims to improve the generalization performance of multiple related tasks by leveraging their relational information [33]. To harness the power of MTL, the learning tasks or a subset of tasks are assumed to be related. For example, a task to detect the positivity of COVID-19 is related to the task designed to assess the severity of the infection. Given the nature of different tasks (imbalance of task difficulty), tasks such as the assessment of the severity can be more difficult to learn than a task to detect infection. Depending on how the hidden layers are shared, there are two different types of MTL approaches, i.e. hard and soft parameter sharing [34]. In the hard parameter sharing, which is more common and capable to greatly reduce the risk of overfitting, hidden layers were shared between all tasks, and each task has its own task-specific output layer(s) [35]. In comparison, the soft parameter sharing approach comes with regularizations to reduce the distance between different task models (each task has its own model and parameters) [34]. Both the two MTL approaches prone to unnecessarily emphasize on easier tasks which can lead to convergence problems for difficult tasks [36]. As a result, specific tasks may dominate the entire learning procedure. Different strategies have been proposed to tackle the problem, for example, Guo et al. introduced adaptive weight adjustment to automatically prioritize more difficult tasks; Liu et al. designed a task-specific network by utilizing an attention module to capture task-related features [37]; Kendall et al. adopted homoscedastic uncertainty (task-dependent uncertainty) as a basis for weighting losses [38]; Tian et al. recently proposed to use two sets of Eigenfunctions (the common one shared by different tasks and unique ones used in individual tasks) to approximate MTL objective function [39].

## 3. COVID-19 CT Studies

A total of 1,329 chest CT studies were enrolled, which including CT scans and corresponding diagnosis and severity assessment results, provided by Wuhan Union Hospital and Wuhan Liyuan Hospital [21]. Each CT scan is corresponding to a single person in the cohort, and studies without chest CT scans were not included. 761 patients were confirmed as COVID-19 positive by nucleic acid test (COVID-NAT). 998 studies were diagnosed as COVID-19 by radiologists using chest CT images (COVID-CT). 237 studies diagnosed as COVID-19 using CT scans but yet confirmed by the nucleic acid test were regarded as "suspected" cases. 331 COVID-19 negative cases were served as "control". The severities of the COVID-19 patients were assessed by physicians based on the infection, symptoms, disease progression, and patient conditions, which can be categorized into control/suspected (type I), mild/regular (type II), and severe/critically ill (type III) (COVID-Severity). The cohort was split arbitrarily into training/cross-validation (70%, *n* = 930) and testing dataset (30%, *n* = 399). The split was stratified by COVID-NAT. The two datasets have similar class distributions. A summary of the study distribution and training/cross-validation vs testing split is shown in Table 1.

**Table 1.** Summary of patient studies and training/cross-validation and testing split.

| Features | Train/Cross-Val (n=930) | Testing (n=399) |
|---|---|---|
| Male/Female | 465/465 | 214/185 |
| Age (mean ± std) | 54.67±16.81 | 53.81±17.80 |
| COVID-CT | | |
|   Positive | 701 | 297 |
|   Negative | 229 | 102 |
| COVID-NAT | | |
|   Positive | 533 | 228 |
|   Negative | 397 | 171 |
| COVID-Severity | | |
|   Control/Suspected | 397 | 171 |
|   Mild/Regular | 398 | 164 |
|   Severe/Critically ill | 135 | 64 |
| Data Source | | |
|   Wuhan Union Hospital | 669 | 290 |
|   Wuhan Liyuan Hospital | 261 | 109 |

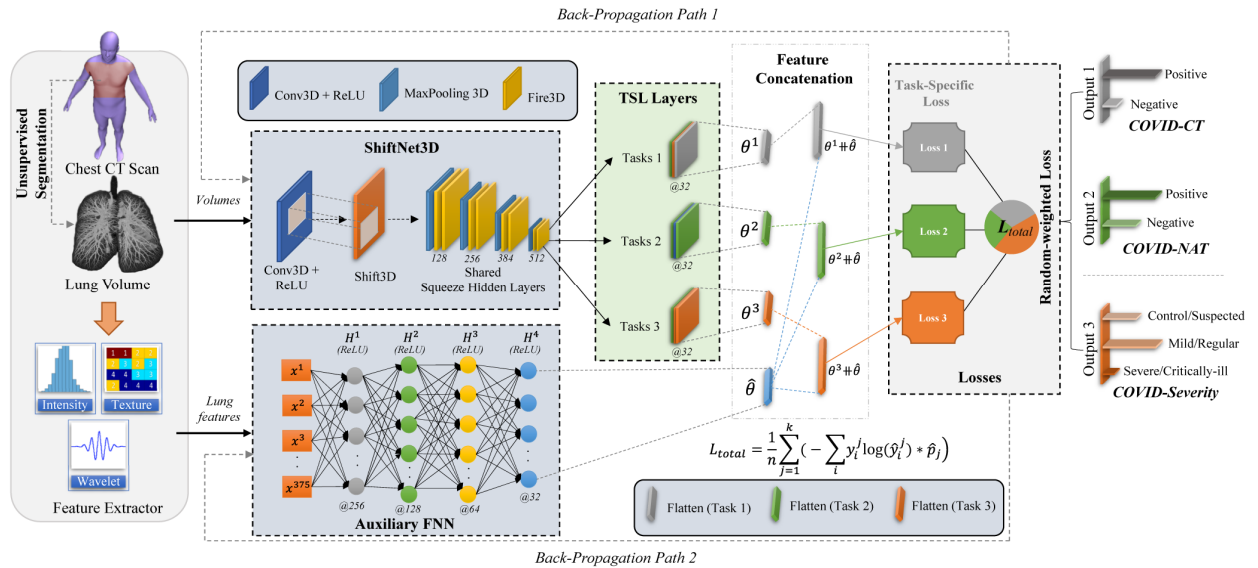

**Fig. 1.** Overview of COVID-19 Multitask Learning Framework.

## 4. Methodology

As illustrated in Figure 1, the proposed multitask learning framework (COVID-MTL) consists of six major components for diagnosis and severity assessment on COVID-19 CT inputs. An unsupervised 3D lung segmentation module is first used to extract lung volumes from chest CT scans. Then, a feature extractor is to obtain high-throughput CT features including intensity, texture, and wavelet features from the segmented lung volumes. Next, the segmented lung volumes and extracted lung features are fed into a ShiftNet3D and a feed-forward neural network (FNN), respectively, to leverage both raw CT data as well as high-throughput imaging features. A hard parameter sharing approach is adopted to construct the MTL model. As the major component of COVID-MTL, ShiftNet3D includes a Shift3D layer to boost network performance through introducing shifting variance on low-level feature representations of the volumetric inputs and 8 consecutive 3D Fire modules (the backbone of SqueezeNet [32]) are used as shared hidden layers between all tasks. To learn task-specific representations, each task has its own output layer (TSL layers) and loss function. High-level feature representations obtained from CT imaging features through the auxiliary FNN (AFNN) are concatenated with each TSL layer for performance enhancement. As such, there are two backpropagation paths in the overall MTL network. Last, a random-weighted loss function is attached to calculate the combined task loss so that different COVID-19 tasks can be trained simultaneously by using weighted total loss as guidance. Key modules of COVID-MTL will be disclosed in section 4.1 – 4.3.

*4.1 Unsupervised Lung Segmentation and High-throughput Lung Feature Extraction*

To avoid the heavy burden on manual labels on lung volumes from chest CT scans during the pandemic, an unsupervised lung segmentation algorithm is more favorable and more viable for mass studies of COVID-19 compared to learning-based methods. However, the widely-adopted unsupervised lung segmentation algorithm [27], which is based on intensity and region connection, is not able to properly handle white lung areas (e.g. GGO) in COVID-19 CT scans. To address this problem, we propose an active contour-based algorithm to refine the initial segmentation results produced by the classical method. The inflated contours of the initially segmented lungs are used as seeds for the refinements. The energy-minimizing refinement method evolves using the given seeds and stops at the boundary of the respective lungs. Thus, avoided inherited under-segmentation defects of thresholding-based methods when dealing with more complicated COVID-19 lung regions. The corresponding pseudocode is summarized in Algorithm 1.

Given the limited engagement of radiologists during the outbreak, over a thousand (1056) instead of hundreds of thousands of CT slices were arbitrarily selected from the cohort (which covered patients with various states) and manually delineated as ground-truth under a senior radiologist's supervision for performance comparison. Segmentation masks produced by classical and proposed methods were compared with ground-truth. The segmentation performance was measured by standard metrics, including Dice similarity coefficient, Jaccard index, Matthews correlation coefficient (MCC), and precision, which are defined as:

$$Dice = \frac{2TP}{2TP + FP + FN}$$

$$Jaccard = \frac{Dice}{2 - Dice}$$

$$MCC = \frac{TP*TN - FP*FN}{\sqrt{(TP+FN)*(TP+FP)*(TN+FP)*(TN+FN)}}$$

$$Precision = \frac{TP}{TP + FP}$$

where $TP$: true positive; $TN$: true negative; $FP$: false positive; $FN$: false negative.

After the lung volumes were automatically segmented by the proposed unsupervised algorithm, a total of 375 high-throughput lung features, which including First Order Statistics, Gray Level Cooccurrence Matrix (GLCM), Gray Level Run Length Matrix (GLRLM), Gray Level Size Zone Matrix (GLSZM), and Wavelet features, were extracted from corresponding lung volumes for the cohort study. To extract wavelet features, Coiflets 1 (coif1) low- and high-pass filters were applied in each of the three dimensions which yield 8 sub-bands (or decompositions). GLCM and GLRLM features were then derived from each sub-band.

```
Algorithm 1: Unsupervised Segmentation Refinement
Input: initial segmented CT scan: CT'
Output: output refinement: out
masks = []
foreach slice ∈ CT' do
    /* Get initial results                    */
    rprobs = regionprops(slice)
    lungs = getlungs(rprobs)
    mask = emptymask(slice, 0)
    /* Process each lung                      */
    foreach lung ∈ lungs do
        /* Obtain refinement seeds            */
        bins = morphoclosing(lung, 10)
        seeds = findcontour(bins)
        /* Refine initial results             */
        snake = activecontour(lung, seeds)
        /* Save refinement results            */
        mask̂ = emptymask(lung, 0)
        mask̂[snake.X, snake.Y] = 1
        mask̂ = fillholes(mask̂)
        mask = mask | mask̂
    /* Save refined 2D mask                   */
    masks.append(mask)
/* Obtain final results with 3D lung mask     */
masks = to3d(masks)
out = CT[masks]
return out
```

**Algorithm. 1.** Pseudocode code of unsupervised refinement method.

*4.2 Shift3D*

Chest CT is more sensitive for early diagnosis of COVID-19 than NAT and other imaging modalities. However, the existing COVID-19 diagnosis based on CT slice unavoidably needs expert involvements. To automate the diagnostic workup, it is necessary to directly process the volumetric radiographic images. 3D CNNs are specifically designed to harness volumetric inputs, but they are notoriously difficult to train, e.g. slow convergence, extremely high memory and computational costs. Therefore, more efficient structures such as 3D SqueezeNet were instead used. However, the lightweight 3D CNNs come with accuracy degradation and still prone to overfitting and suffering from slow convergence. To address those problems and make them more feasible in practice, here we propose a 3D real-time augmentation method, named Shift3D, which introduces space variances through randomly shifting low-level feature representations of the volumetric inputs in three dimensions (or 6 directions). The rationale of this setting is based on our observation that the geographical location of human organs in CT scans varies from one case to another, and even for different scans of the same patient (people lying down on a CT bed without exactly the same positions). Such space variances may affect the network performance and thus worth to be dealt with. In comparison to traditional augmentation methods, Shift3D directly operate on different levels of feature representations (3D feature maps) instead of original volumetric inputs and leverage GPU computing power by being implemented as a neural network layer. Related studies with 2D CNNs suggests that nonlinearly augment 2D feature maps can achieve better performance compared to input augmentation [40, 41]. 3D scaling may also introduce space variances, but it is more computationally expensive.

The pseudocode of Shift3D is illustrated in Algorithm 2. There are three parameters for Shift3D: max shift percentage $p$ (default is 0.2) decides the maximum percentage of a shift in each of the 6 directions (in compared to the size of the corresponding dimension); elements will be re-introduced at the first position if they are shifted beyond the last position, and the *ispadding* and *padding$_v$* are used to fill re-introduced elements with specific numbers. The usage of Shift3D is flexible, for example, one can lower the frequency for calling Shift3D in a wrapped network layer to reduce shifting chance and processing power; it can also combine with existing augmentation methods to further boost network performance. Similar to other augmentation algorithms, it is not recommended to use Shift3D in the inference stage.

```
Algorithm 2: Shift3D
Input: input tensor: x, max shift percentage: p, shift
       padding: ispadding, padding value: padding_v
Output: output tensor out
/* Shift dimension and direction              */
dim = randint(0, 2)
forward = −1 if randint(0, 1) > 0 else 1
/* The number of lines to shift               */
shifts = randint(0, int(p ∗ x.shape[dim]))
/* Shift in specified dimension and direction */
out = torch.roll(x, shifts = forward ∗ shifts, dims = dim)
/* Perform padding if specified               */
if ispadding then
   ⌊ out = padding(out, forward < 0, padding_v)
return out
```

**Algorithm. 2.** Pseudocode implementation of Shift3D with PyTorch.

*4.3   Random-weighted Multitask Loss*

Difficult tasks like severity assessment of COVID-19 may induce higher losses compared to easier tasks (e.g. diagnosis) and thus more vulnerable in the MTL learning procedure, e.g. slow convergence and lower learning priority. Inspired by task-dependent uncertainty loss proposed by Kendall et al. [38], where uncertainty weights were learned by tuning log variances, here we propose a random-weighted loss function, which randomly assigns learning weights to different tasks during each iteration of the joint training, to prevent the learning procedure being dominated by any specific tasks. The random weights are drawn from the Dirichlet distribution since it can generate probability distributions that satisfy our needs: 1) sum of all task weights (probabilities) is equal to 1; 2) it allows us to control the concentration of a generated weight distribution (discuss later). Such a random-weighted setting is based on the probability theory that each of the $k$ tasks has $\sim 1/k$ chance to be prioritized if the number of iterations in joint training is large. Therefore, vulnerable tasks still have a sufficient chance to be prioritized and trained.

The Dirichlet distribution uses a probability density function that defined as:

$$\mathrm{p}(P,\alpha) = \frac{1}{B(\alpha)} \prod_{i=1}^{k} p_i^{\alpha_i - 1} \quad (1)$$

where, $K \geq 2$, $K$ is the number of learning tasks; $P = (p_1...p_n)$, $\sum_1^k p_i = 1; p_i \geq 0$, $p_i$ is the weight of learning task $i$; $B(\alpha)$ is the normalization constant and $\alpha = (\alpha_1...\alpha_k)$, which can be expressed as a gamma function:

$$B(\alpha) = \frac{\prod_{i=1}^{k} \Gamma(\alpha_i)}{\Gamma(\sum_{i=1}^{k} \alpha_i)} \quad (2)$$

Since the objective function for each task is a cross-entropy loss, which is defined as:

$$L(\hat{y}, y) = -\sum_i y_i \log(\hat{y}_i) \quad (3)$$

The total loss function of a MTL model with random-weighted loss, therefore, can be calculated as:

$$L_{total}(\hat{y}^1...\hat{y}^k, y^1...y^k) = \sum_{j=1}^{k} (-\sum_i y_i^j \log(\hat{y}_i^j) * p_j) \quad (4)$$

For the special case when $K=2$, the weights for the two learning tasks can be simply decided as:

$$L_{total}(\hat{y}^1, \hat{y}^2; y^1, y^2) = (-\sum_i y_i^1 \log(\hat{y}_i^1) * p) + (-\sum_i y_i^2 \log(\hat{y}_i^2) * (1-p)) \quad (5)$$

where $p$ can be drawn from either Dirichlet or others like Uniform distribution when there are only two tasks.

One can draw random weights $n$ times and average the results to avoid potential heavy fluctuation of a single Dirichlet draw, thus the total loss function can be finally modeled as:

$$L_{total}(\hat{y}^1...\hat{y}^k, y^1...y^k) = \frac{1}{n} \sum_{j=1}^{k} (-\sum_i y_i^j \log(\hat{y}_i^j) * \hat{p}_j) \quad (6)$$

while $\hat{p}_j$ is the accumulation weights of $n$ draws for the task $j$. $n$ can be seen as a hyperparameter which controls weight difference (or concentration) between $K$ tasks, the larger the $n$ the less difference among tasks' weights (evenly distributed), therefore, the random-weighted loss will be downgraded to simple mean loss if $n$ is large enough. In other words, a larger $n$ is preferred if tasks' difficulties are similar, otherwise, a smaller $n$ is better for tasks that have distinct difficulties.

As indicated in equation 6, instead of tuning normalization constant $\alpha$ (a vector of concentration parameters), we fixed $\alpha = (1...1)$ which made Dirichlet distribution generate high concentrated probabilities by default, and then we introduced a single integer parameter $n$ to control the concentration (which is much easier to tune), and the degree for prioritizing vulnerable task(s).

*4.4 Pattern Analysis of High-throughput Lung Features and their Correlation with COVID-19*

Because of the deep features extracted by the neural network lack interpretability. To decipher the correlation between CT images and COVID-19, high-throughput imaging features extracted from CT lung volumes were instead analyzed. The feature studies were stratified into different groups based on the positivity and severity of COVID-19. Because the stratified groups may not have equal feature variances and equal sample sizes, Welch's ANOVA was therefore used to test differences between the group means. The top imaging features that are significantly related to COVID-19 were then identified based on Welch's results. The patterns of high-throughput features and their correlation with COVID-19 infection and severity as well as routine clinical parameters like gender and age can be further analyzed with clustering heatmap, where imaging features were first scaled by z-score and then hierarchically clustered based on the distance of the Pearson correlation coefficient.

## 5. Experiments and Results

*5.1. Experiment settings*

A total of six machine learning and deep learning models were used in this study for diagnosis and severity assessment of COVID-19. All of the chest CT scans were resampled into the spacing of 1mm $x$ 1mm $x$ 1mm for lung segmentation. High-throughput features extracted from lung CT volumes were used to train Random Forests (RF) and LightGBM (LGBM) models. In each of the two machine learning models, 1,000 estimators (decision trees) were utilized. A learning rate of 0.01 was used to train the LGBM model. The number of feature maps of ShiftNet3D is ranging from 64 to 512; in comparison, the AFNN branch contains 256, 128, 64, and 32 nodes in its four hidden layers. To train deep learning models, i.e. ResNet3D (ResNet34 structure), SqueezeNet3D, ShiftNet3D, and COVID-MTL, standard preprocessing and augmentation procedures were utilized, which including normalization, random rotation, flip, and crop (size of 200 $x$ 250 $x$ 250 pixels covering major lung regions). As for the training parameters, stochastic gradient descent (SGD) with Nesterov momentum of 0.9, weight decay of 5E-5, 80 epochs, and a batch size of 10 was used for all CNN models. A cosine learning rate scheduler [39, 40] was utilized and the learning rate was started with 0.005 and gradually declined to a minimum of 5E-5. He-normalization was adopted to initialize network weights. The number of draws ($n$) for Dirichlet distribution in each training iteration is 2. The abovementioned hyperparameters were derived from the training/cross-validation dataset. More complicated 3D ResNet models, like the 3D version of ResNet-50, unable to be trained due to the limitation of GPU memory capacity. All CNN models were trained on two Nvidia RTX 2080Ti GPUs using the consistent settings. COVID-MTL can be trained with or without high-throughput inputs. The last epoch testing performances, which measured by precision, recall, F1 score, accuracy, and area under the curve (AUC), were reported.

*5.2. Unsupervised Lung Segmentation on COVID-19 Chest CT Scans*

**Table 2.** Performance metrics of lung segmentation on COVID-19 CT scans.

| Method | Dice | Jaccard | MCC | Precision |
|---|---|---|---|---|
| Classical | 0.971 | 0.945 | 0.966 | 0.961 |
| Proposed | 0.982 | 0.966 | 0.979 | 0.981 |

As shown in Table 2, both the classical method [27] and the proposed algorithm achieved high segmentation performance. The proposed refinement method consistently improved the state-of-the-art and the benefits are expected coming from the improvements of under-segmentation in white lung areas, as indicated in Figure 2.

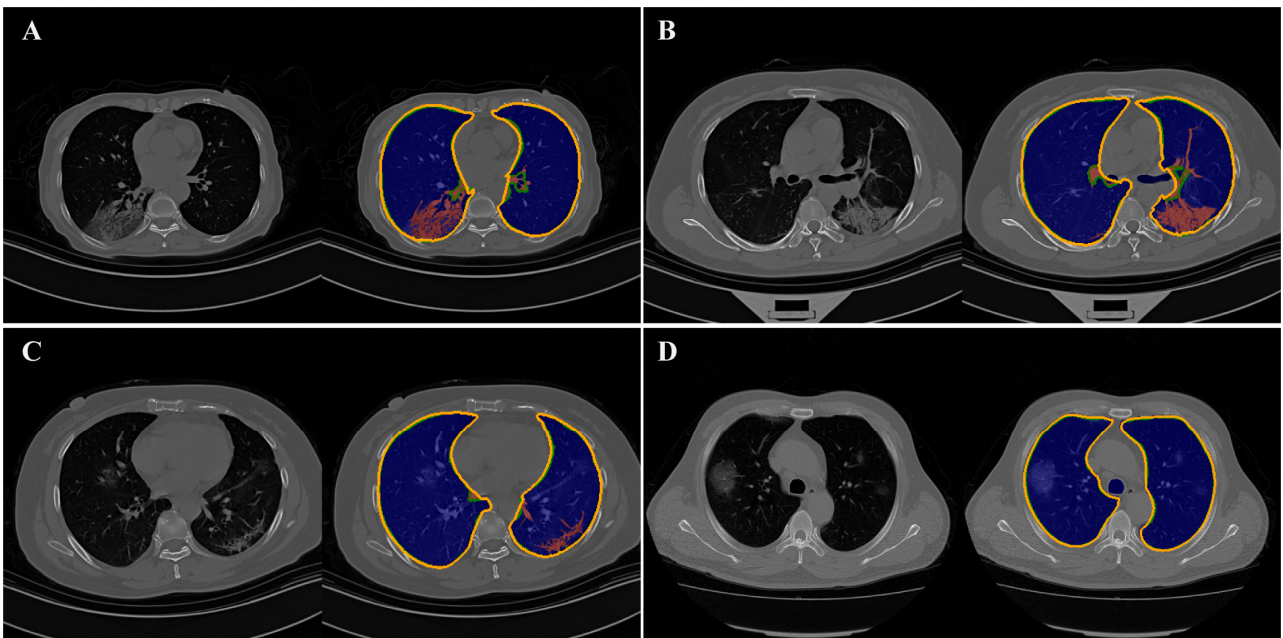

**Fig. 2.** Sample lung segmentation on COVID-19 CT scans. Blue region: segmented by classical method; red regions contains tissue structures like GGO that crucial for COVID-19 diagnosis; orange contour: refinement results; green contour: ground-truth.

The sample result in Figure 2 illustrates that the proposed refinement algorithm can accurately detect white lung areas from COVID-19 CT studies in different scenarios, especially on more challenging cases as illustrated in Figure 2A-C, whereas, the classical method failed to detect crucial COVID-19 diagnostic structures (red region, Figure 2A-C). However, the classical method does not always produce negative outcomes, it can detect ongoing GGOs in some of the cases where mild to moderate white appearance mostly occurred within but not on the edge of the lung (Figure 2D).

*5.3. Experimental Results of COVID-19 Diagnosis*

The six types of models were trained on the full training/cross-validation dataset and then tested on another 399 CT studies (Table 1). Corresponding detection results were assessed against radiology (COVID-CT) and nucleic acid tests (COVID-NAT).

As shown in Figure 3 and Table 3, the two popular machine learning models (RF and LGBM) achieved similar detection performance, i.e. AUCs of 0.913/0.921 and accuracies of 86.47%/86.47% against radiology, and AUCs of 0.819/0.803 and accuracies of 73.93%/76.19% against SARS-CoV-2 nucleic acid tests, using high-throughput lung features.

When directly utilizing 3D CT lung volumes as inputs, 3D CNN models, especially ShiftNet3D, yielded higher performance in comparison to RF and LGBM. Compared to the other two 3D CNN models, ShiftNet3D achieved around 4-5% higher accuracy and AUC performance for detection of COVID-19 against nucleic acid tests (Table 3), suggesting the introduction of Shift3D can induce a performance boost for existing 3D CNNs on the more challenging COVID-19 learning task.

After the adoption of a random-weighted loss function, the COVID-MTL is capable of simultaneously detecting COVID-19 against both radiology (AUC of 0.939, the accuracy of 90.23%) and nucleic acid tests (AUC of 0.846, the accuracy of 79.20%) (Figure 3 and Table 3). In comparison to single-task models mentioned previously, the MTL approach achieved similar capability for detection against radiology but significant higher performance for diagnosis against nucleic acid tests, i.e. over 3% accuracy promotion compared to ShiftNet3D, and around 7-9% compared to ResNet3D and SqueezeNet3D (Table 3), suggesting the effectiveness of joint learning. Even higher performance was achieved when using CT lung volumes and high-throughput imaging features as parallel inputs. COVID-MTL models equipped with Shift3D consistently outperform model that without Shift3D. It is worth noting that the training and inference time of the COVID-MTL models is significantly reduced (three times less) in comparison to conventional 3D CNN models since the latter needs to be trained and predicted individually for each task.

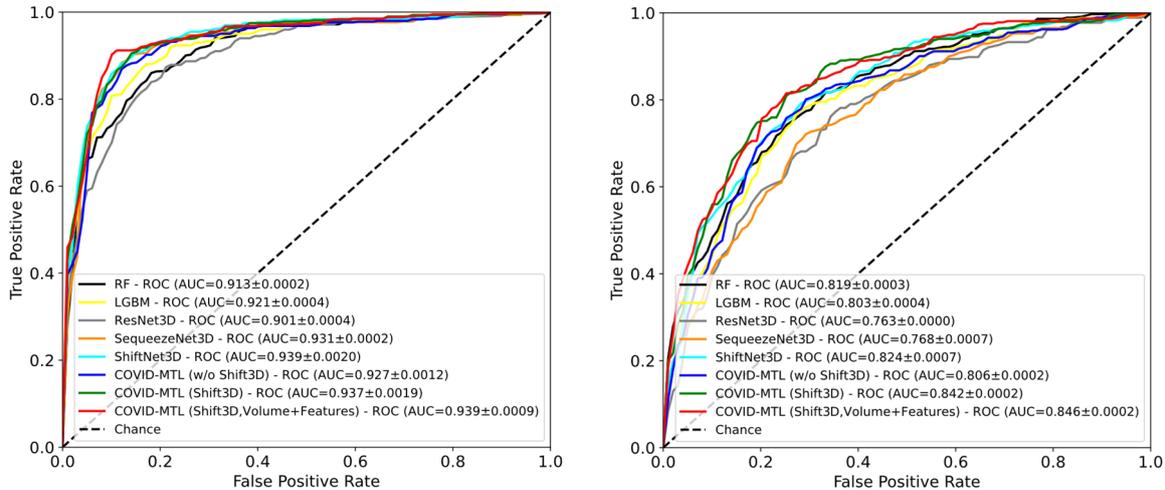

**Fig. 3.** ROC/AUCs of machine learning and deep learning models for detection of COVID-19 against radiologists (left) and SARS-CoV-2 nucleic acid test (right).

**Table 3. Performance metrics of machine learning and deep learning models for COVID-19 diagnosis.**

| Model | | Input(s) | COVID-19 against Radiology | | | | | COVID-19 against Nucleic Acid Test | | | | |
|---|---|---|---|---|---|---|---|---|---|---|---|---|
| | | | Prec. | Rec. | F1 | Acc. | AUC | Prec. | Rec. | F1 | Acc. | AUC |
| RF | | CT Features | 0.861 | 0.865 | 0.862 | 86.47% | 0.913 | 0.739 | 0.739 | 0.739 | 73.93% | 0.819 |
| LGBM | | CT Features | 0.862 | 0.865 | 0.863 | 86.47% | 0.921 | 0.761 | 0.762 | 0.761 | 76.19% | 0.803 |
| ResNet3D | | CT Volume | 0.841 | 0.840 | 0.840 | 83.96% | 0.901 | 0.723 | 0.724 | 0.724 | 72.43% | 0.763 |
| SqueezeNet3D | | CT Volume | 0.897 | 0.885 | 0.888 | 88.47% | 0.931 | 0.713 | 0.707 | 0.708 | 70.68% | 0.768 |
| ShiftNet3D | | CT Volume | 0.896 | 0.887 | 0.890 | 88.72% | **0.939** | 0.762 | 0.762 | 0.762 | 76.19% | 0.824 |
| COVID-MTL | w/o. Shift3D | CT Volume | 0.891 | 0.877 | 0.881 | 87.72% | 0.927 | 0.760 | 0.757 | 0.758 | 75.69% | 0.806 |
| COVID-MTL | Shift3D | CT Volume | 0.891 | 0.882 | 0.885 | 88.22% | 0.937 | **0.796** | **0.794** | 0.791 | **79.45%** | 0.842 |
| COVID-MTL | Shift3D | CT Volume, CT Features | **0.912** | **0.902** | **0.905** | **90.23%** | 0.939 | 0.791 | 0.792 | **0.792** | 79.20% | **0.846** |

Figure 4 illustrates the total loss of COVID-MTL models (with CT volumetric inputs) for COVID-19 detection with and w/o using the Shift3D. The losses of different tasks were randomly weighted and summed during each training iteration, and the corresponding total test loss fluctuated as expected in the earlier learning stage. With the help of the Shift3D, the multitask learning model converged faster, and the fluctuation of the total loss was also alleviated (Figure 4).

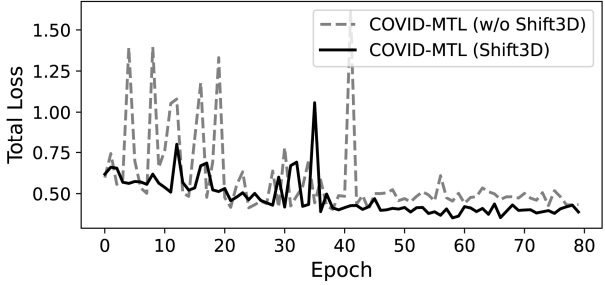

**Fig. 4.** Total loss comparison of COVID-MTL for detecting COVID-19 against both radiology and SARS-CoV-2 nucleic acid test under using and w/o using Shift3D.

### 5.4. Experimental Results of COVID-19 Severity Assessment

As shown in Table 4, except ResNet3D, the deep learning models achieved consistently higher performance compared to RF and LGBM for COVID-19 severity assessment. ShiftNet3D yielded similar performance compared to its backbone model (SqueezeNet3D). In comparison, the COVID-MTL achieved a slight performance boost with an AUC of 0.800 ± 0.020 and an accuracy of 66.67%.

**Table 4.** Performance metrics of machine learning and deep learning models for severity assessment of COVID-19.

| Model | Pre. | Recall | F1 | Acc. | AUC |
|---|---|---|---|---|---|
| RF | 0.628 | 0.639 | 0.624 | 63.91% | 0.791 |
| LGBM | 0.630 | 0.647 | 0.632 | 64.66% | 0.784 |
| ResNet3D | 0.546 | 0.556 | 0.549 | 55.64% | 0.737 |
| SqueezeNet3D | 0.655 | 0.659 | **0.653** | 65.91% | 0.794 |
| ShiftNet3D | 0.655 | 0.659 | **0.653** | 65.91% | 0.794 |
| COVID-MTL | **0.666** | 0.667 | 0.649 | 66.67% | 0.800 |
| COVID-MTL (Transfer) | 0.647 | **0.669** | 0.632 | **66.92%** | **0.813** |

Other than training three tasks together, it is reasonable to assume that the MTL model trained for the two diagnosis tasks can be reused for severity assessment since the control/suspected cases can be inferred from the positivity of the CT diagnosis and nucleic acid tests. To validate the hypothesis, the COVID-MTL model trained for the two diagnosis tasks were repurposed using transfer learning for severity assessment. The task-specific output and classification layers of COVID-MTL were replaced with fully-connected layers, the pretrained convolutional layers were frozen and the reused model was then trained for additional 50 epochs. As a result, the transfer learning model achieved an AUC of 0.813 ± 0.021 for classifying control/suspected (AUC of 0.841), mild/regular (AUC of 0.808), and severe/critically-ill (AUC of 0.789) cases, which is a slight boost for original COVID-MTL model. Interestingly, the 3D implementation of ResNet (ResNet34 in this work), which was based on [31], achieved the lowest performance on severity assessment. More complicated 3D ResNet models unable to be loaded under the present settings due to the GPU memory limitation. Even deeper ResNet structures could be converted to 3D versions and explored when enough GPU resources available.

**Table 5.** Welch's ANOVA test of top CT lung features between COVID-19 positive and negative cases, and among different severity groups.

| | Robust Tests of Equality of Means | | | | | | | | |
|---|---|---|---|---|---|---|---|---|---|
| | COVID-19 against Radiologists (df1=1) | | | COVID-19 against Nucleic Acid Test (df1=1) | | | COVID-19 Severity (df1=2) | | |
| High-throughput Lung Features | Stat. | df2 | Sig. | Stat. | df2 | Sig. | Stat. | df2 | Sig. |
| HLL_glcm_ClusterProminence | 43.06 | 340.89 | <0.001 | 61.80 | 604.97 | <0.001 | 30.86 | 472.32 | <0.001 |
| LHL_glcm_Idmn | 30.06 | 530.47 | <0.001 | 30.43 | 1198.09 | <0.001 | 18.10 | 540.37 | <0.001 |
| Maximum | 80.41 | 441.28 | <0.001 | 57.10 | 956.95 | <0.001 | 29.73 | 542.92 | <0.001 |
| Energy | 2.49 | 451.09 | 0.115 | <0.01 | 1096.54 | 0.986 | 10.52 | 514.79 | <0.001 |
| LLL_glcm_Imc1 | 409.21 | 668.77 | <0.001 | 167.23 | 1208.49 | <0.001 | 112.46 | 541.73 | <0.001 |
| HLL_glcm_Correlation | 29.58 | 512.00 | <0.001 | 1.65 | 1119.28 | 0.199 | 1.19 | 591.13 | 0.306 |
| LLH_glcm_Correlation | 3.89 | 604.49 | 0.049 | 7.79 | 1321.56 | 0.005 | 6.45 | 539.50 | 0.002 |
| LHH_glcm_ClusterShade | 77.59 | 393.65 | <0.001 | 114.05 | 817.28 | <0.001 | 75.90 | 529.59 | <0.001 |
| *LongRunLowGrayLevelEmphasis | 54.96 | 689.20 | <0.001 | 68.80 | 1306.63 | <0.001 | 37.82 | 489.17 | <0.001 |
| HLH_glcm_ClusterShade | 70.23 | 402.28 | <0.001 | 127.16 | 868.70 | <0.001 | 82.64 | 522.13 | <0.001 |
| Idn | 13.55 | 530.18 | <0.001 | 7.65 | 1158.71 | 0.006 | 6.35 | 546.89 | 0.002 |
| LLH_glcm_ClusterShade | 2.96 | 489.89 | 0.086 | 0.05 | 1191.24 | 0.832 | 0.21 | 595.76 | 0.807 |
| LargeAreaHighGrayLevelEmphasis | 1.07 | 537.72 | 0.300 | <0.01 | 1311.65 | 0.993 | 4.31 | 515.67 | 0.014 |
| *ShortRunHighGrayLevelEmphasis | 44.66 | 360.21 | <0.001 | 50.89 | 702.63 | <0.001 | 27.48 | 475.40 | <0.001 |
| Idmn | 8.63 | 534.83 | 0.003 | 0.72 | 1146.50 | 0.395 | 1.32 | 549.05 | 0.268 |
| GrayLevelVariance | 84.66 | 442.59 | <0.001 | 124.43 | 1030.40 | <0.001 | 63.26 | 582.43 | <0.001 |
| HHH_glcm_ClusterShade | 62.03 | 377.48 | <0.001 | 108.32 | 753.05 | <0.001 | 64.82 | 534.31 | <0.001 |
| LLL_glcm_Imc2 | 346.27 | 630.78 | <0.001 | 107.75 | 1178.31 | <0.001 | 93.34 | 546.17 | <0.001 |
| LLL_glrlm_RunEntropy | 47.48 | 548.11 | <0.001 | 2.72 | 1237.29 | 0.100 | 43.13 | 515.62 | <0.001 |
| LLH_glcm_ClusterProminence | 29.54 | 429.01 | <0.001 | 25.00 | 1032.16 | <0.001 | 14.92 | 598.61 | <0.001 |
| *DifferenceVariance | 129.02 | 383.67 | <0.001 | 143.71 | 792.11 | <0.001 | 71.98 | 567.16 | <0.001 |
| Imc2 | 279.23 | 651.74 | <0.001 | 48.15 | 1178.92 | <0.001 | 54.21 | 530.26 | <0.001 |
| LLL_glcm_Correlation | 171.47 | 512.35 | <0.001 | 173.88 | 1138.84 | <0.001 | 89.70 | 539.75 | <0.001 |
| Imc1 | 253.18 | 673.94 | <0.001 | 51.66 | 1208.52 | <0.001 | 42.87 | 533.73 | <0.001 |

Stat. *Asymptotically F distributed*; *LongRunLowGrayLevelEmphasis: *HLH_glrlm_LongRunLowGrayLevelEmphasis*;
*ShortRunHighGrayLevelEmphasis: *LHH_glrlm_ShortRunHighGrayLevelEmphasis*; *DifferenceVariance: *HLL_glcm_DifferenceVariance*

*5.5. Pattern Analysis of High-throughput Lung Features and their Correlation with COVID-19*

For each COVID-19 task, feature importance was generated after training the machine learning model (LGBM in this work). The top 10 most important imaging features were selected for each task. After removing repeated features, 24 high-throughput lung features remained for correlation analyses. As mentioned in the methods section, the 24 imaging features under different COVID-19 states were analyzed using Welch's ANOVA test, which then identified 16 image features that significantly related to COVID-19 positivity and severity ($P < 0.001$ for all; gray items in Table 5). 10 significant features were further scaled to the same value range and were box-plotted for better illustration (Figure 5).

In other words, those identified features are significantly different between COVID-19 positive and negative cases against both radiology ($P < 0.001$; second column, Table 5; first row of Figure 5) and SARS-CoV-2 nucleic acid test ($P < 0.001$; third column, Table 5; second row of Figure 5). Similar results were observed in the analysis of COVID-19 severity, where significant value differences were found in different severity groups ($P < 0.001$, last column, Table 5), especially when comparing control/suspected and severe/critically ill cases (last row, Figure 5).

As shown in Figure 6, clustering of the top lung imaging features demonstrated that the nuclei acid test results were not always consistent with the diagnosis from radiologists (using CT scans), which is in accordance with the published literature [7, 42]. The inconsistency between the two diagnosis standards may due to the high false-negative rate of NAT. The clustering also shows that COVID-19 infection can be found in different age and gender groups (Figure 6). People in older age groups are more vulnerable to be with severe/critically-ill infection (Figure 6) given their immune response is less effective to SARS-CoV-2 compared to their younger counterparts [43-45]. More interestingly, there is a group of uninfected people (mostly the male) whose lung CT features demonstrated a distinct pattern compared with others (middle, Figure 6).

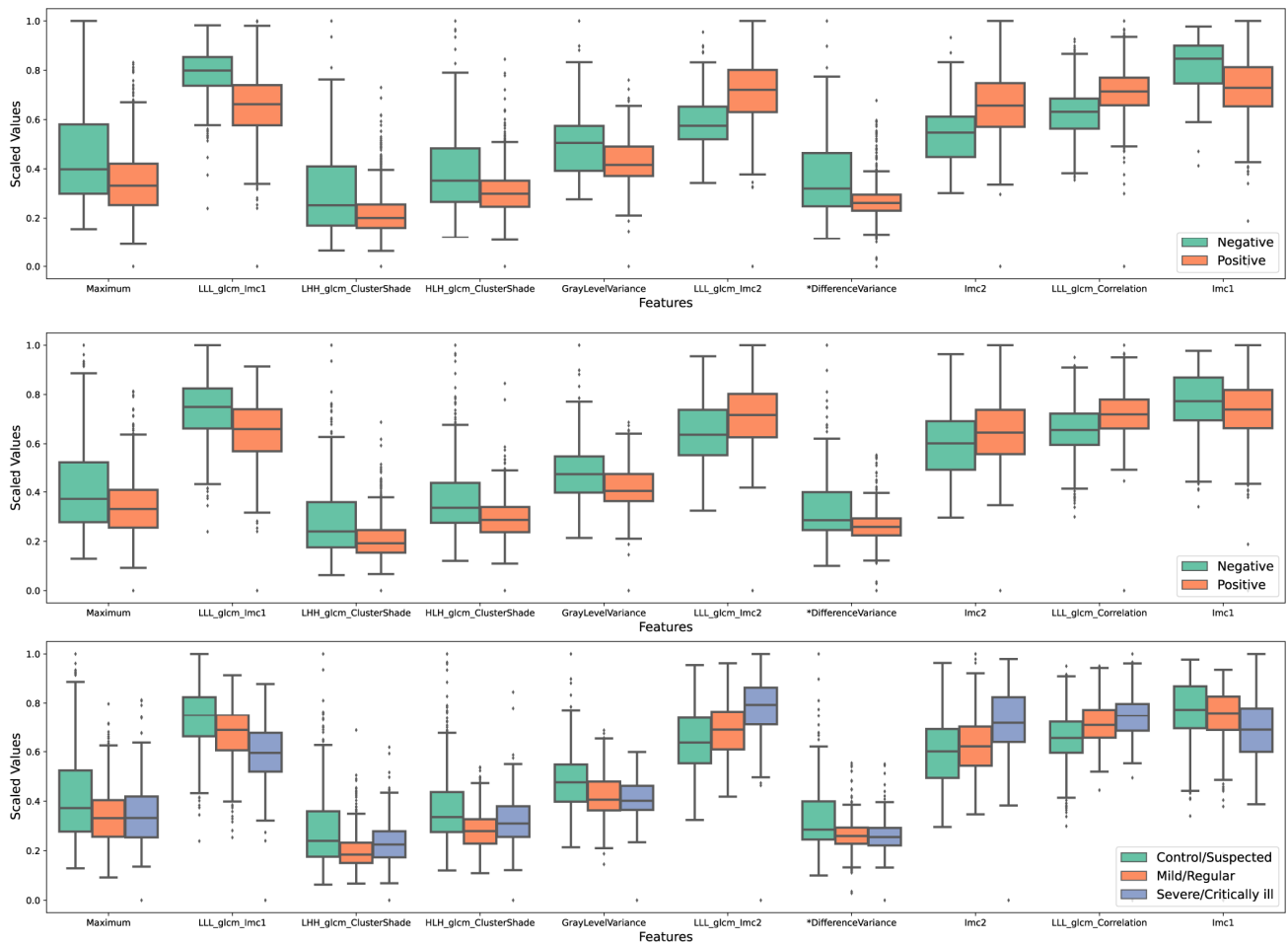

**Fig. 5.** Box plots of CT lung features between COVID-19 positive and negative cases (first row: against radiology; second row: against SARS-CoV-2 nucleic acid test) as well as among different severity groups (last row).

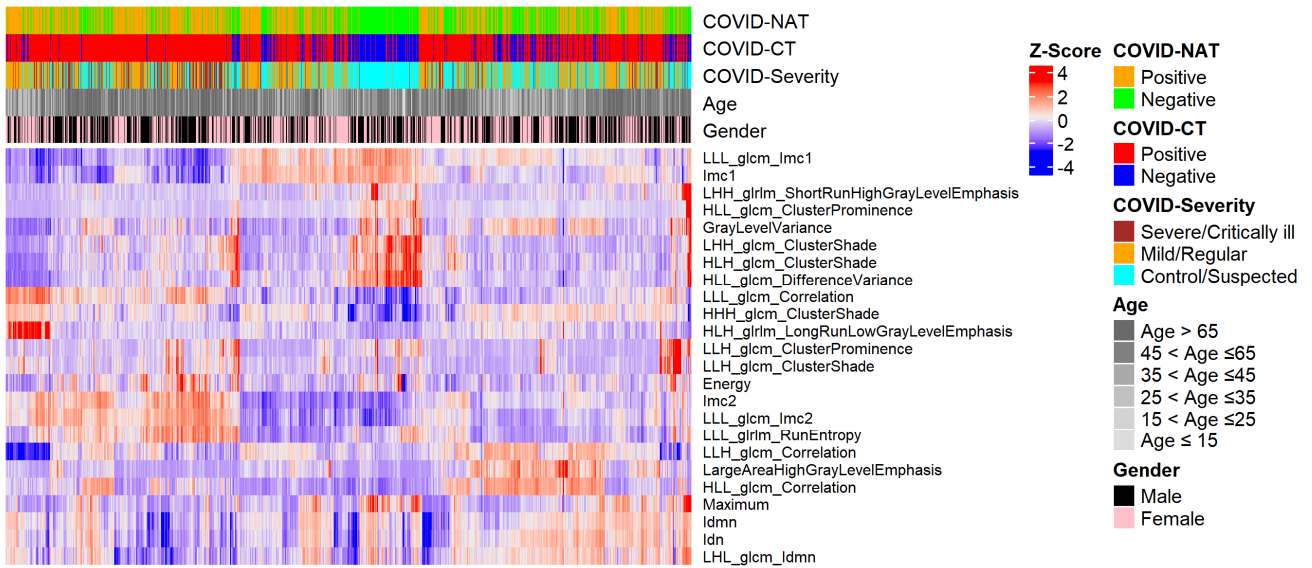

**Fig. 6.** Clustering of top high-throughput lung features that related to COVID-19.

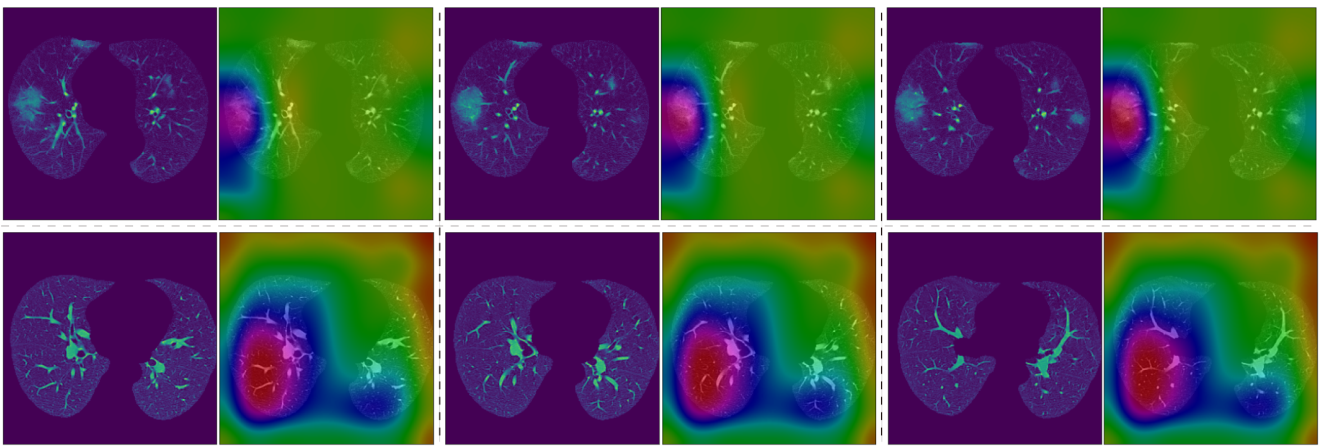

**Fig. 7.** CAM visualization for comparison of discriminative regions captured by COVID-MTL in the diagnosis of COVID-19. Upper panel: lung region and corresponding CAM visualization of an infected case (ongoing infection with severe symptoms, ground-glass opacities exhibited); Bottom panel: lung region and corresponding CAM visualization of a normal case.

## 6. Case study

To decipher the underlying mechanism of COVID-MTL for detection of COVID-19 infection, we obtained 3D feature maps from the last convolutional layer of COVID-MTL when inferencing an infected case (upper panel, Figure 7) and a normal study (bottom panel, Figure 7). The feature maps were then converted to Class Activation Maps (CAMs) and overlaid on the two cases respectively to compare the discriminative regions captured by the MTL model. The comparison shows a distinct discriminative pattern between the two cases, which indicate that the discriminative regions captured from the infected case are focused on lung areas that exhibited ground-glass opacities (red attention color in the upper panel, Figure 7), whereas, large and homogeneous lung tissue regions were covered in the normal case (red attention color in the bottom panel, Figure 7).

A comparison of their lung CT features is also demonstrated in Figure 8, in which some of the features are significantly different between the two cases, including Imc1, HLL_glcm_DifferenceVariance, HLH_glcm_ClusterShade, LHH_glcm_ClusterShade, and LLL_glcm_Imc1.

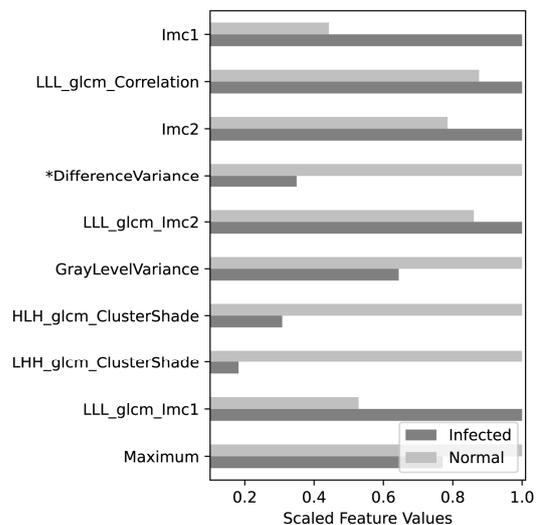

**Fig. 8.** Comparison of top high-throughput lung features between an infected and a normal case. Features are scaled to 0 - 1 for comparison.

## 7. Ablation study

We compared the performance of task-dependent uncertainty loss, random-weighted loss, and mean loss (average of 3 task losses) under COVID-MTL for detection and severity assessment of COVID-19. The result shows our proposed random-weighted multitask loss function achieved faster and stable convergence as well as better performance in comparison to task-dependent uncertainty loss [38] and mean loss (upper panel, Figure 9). COVID-MTL models equipped with Shift3D achieved consistently better performance than models without Shift3D, i.e. faster convergence and better performance have been achieved under all three types of loss functions with random-weighted loss slightly better than the other two methods (bottom panel, Figure 9).

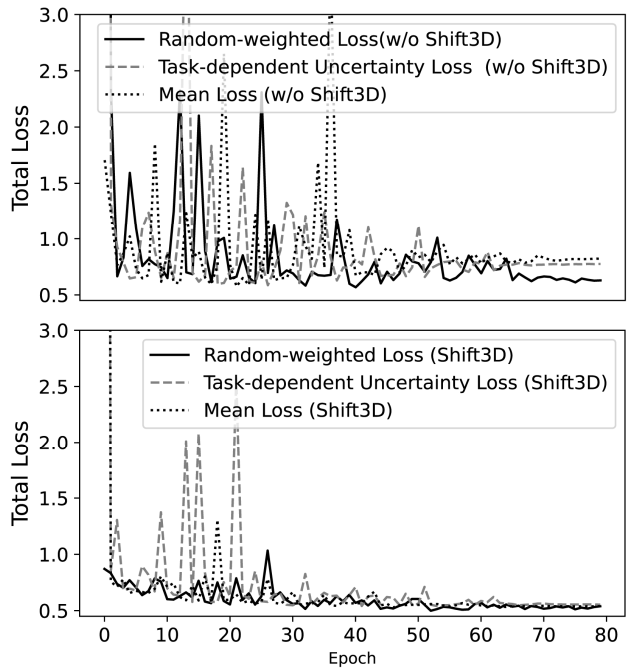

**Fig. 9.** Performance of three different multitask loss functions for detection and severity assessment of COVID-19 when w/o using (up) and using (down) Shift3D.

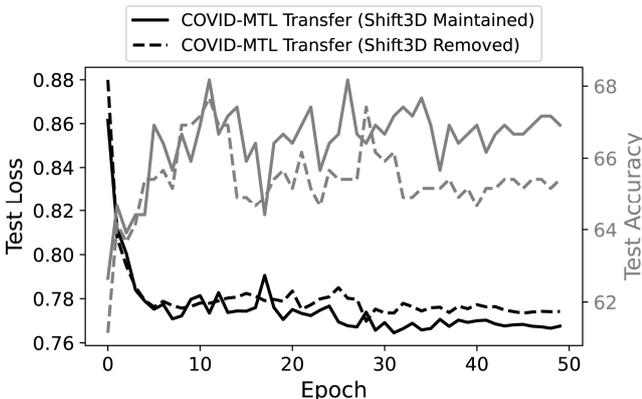

**Fig. 10.** Performance comparison of COVID-MTL transfer learning model for severity assessment of COVID-19 when maintaining (solid lines) and removing (dashed lines) Shift3D.

As we have shown before, the COVID-MTL model trained for the two diagnosis tasks can be repurposed for severity assessment (Table 4), which achieved a slight performance boost (AUC of 0.813 ± 0.021, accuracy of 66.92%, and recall of 0.669) compared to the original MTL model (AUC of 0.800 ± 0.020, accuracy of 66.67%, and recall of 0.667). However, such performance gain was obtained when the Shift3D layer was still enabled during the retraining procedure. Without the utilization of Shift3D, the transfer learning model unable to achieve such performance gain (AUC of 0.810 ± 0.024, accuracy of 65.41%, and recall of 0.654) even it was trained for additional 50 epochs. This finding was also illustrated in Figure 10.

## 8. Discussion and Conclusion

With the dramatic increase of COVID-19 infections in the past few months and the shortage of human resources in clinical practice globally, there is an urgent need for the automated methods to help physicians diagnose and assess the severity of the highly infectious disease. Based on the cross-institutional CT studies, we proposed a multitask learning framework to simultaneously detect and evaluate the severity of COVID-19. Both radiology and NAT have their pros and cons, they are thereby widely adopted as complimentary diagnosis methods in clinical practice since the outbreak. COVID-MTL demonstrated its capabilities to achieve high performance on COVID-19 detection against radiology (computer tomography), meanwhile, the framework can simultaneously infer NAT results which indicated deep features obtained from CT scans may contain richer information for the assessment of COVID-19 than as expected. By combining 3D CNN and auxiliary FNN, different representations of chest CT information, i.e. volumetric lung CT data and high-throughput lung CT features, were propagated and concatenated in the network for evaluation of COVID-19. The fine-grained detection results can be obtained based on the prediction against both radiology and NAT, however, the diagnosis of COVID-19 is more complicated than expected due to prolonged incubation and asymptomatic infection from which biomedical tests and repeated swab samples need to be collected and evaluated periodically to confirm suspected cases.

The proposed Shift3D and random-weighted multitask loss function were experimentally validated to be able to improve the convergence and accuracy of the state-of-the-art methods, especially on more challenging COVID-19 tasks. Shift3D alleviates the overfitting of existing 3D CNNs by introducing space variances (3D displacements) and works well under different loss configurations. In comparison, the random-weighted loss function gives vulnerable tasks a sufficient chance to be prioritized and prevents joint learning procedure from being dominated by specific tasks, which outperformed task-dependent uncertainty loss and linearly combined mean loss. Dirichlet distribution allows us to control the concentration of the generated weight distributions by fixing the concentration parameter vector $\alpha$ and tuning a drawing number $n$ instead, which is quite useful when dealing with imbalanced tasks (in terms of task difficulty). Distributions like the normal ones can also be used when there are only two tasks. However, other probability distributions that satisfy the abovementioned needs may also effective, which deserves further exploration. More studies need to be carried out in the future to validate the performance of the current loss setting on other tasks, especially on 2D CNNs and under the combination of classification and regression tasks.

COVID-MTL works under the utilization of chest CT scan only, it is a self-contained framework and can work independently without human intervention, thus reduce inter and intra-observer variability especially compared to slice-based methods in which the inference can be heavily affected by the inputs (quality of manually labeled CT slices). COVID-MTL is independent of clinical parameters but can be further enhanced by integrating those clinical factors, including biochemical tests. In comparison to single-task solutions, the performance of different tasks, especially the more challenging ones, was boosted under the joint learning, and meanwhile, training and inference time can be significantly reduced given its capability of learning and predicting three different COVID-19 tasks in parallel.

The deep learning model is more like a black box, to decipher the relationship between radiographic images and COVID-19, high-throughput lung features were extracted from COVID-19 CT scans, and top imaging features were identified through statistical analyses. The analyses showed that those lung CT features are significantly ($P < 0.001$) related to COVID-19 positivity and severity. The following case study, which was conducted to decipher the underlying mechanism of COVID-MTL for recognition of COVID-19, ascertained the findings by showing the distinct discriminative patterns that were captured by the neural network from CT images of the infectious and normal case respectively. The analyses of high-throughput lung features and their correlation with COVID-19 may help the community better understand the disease regarding its relevance to radiology

In conclusion, we proposed an end-to-end multitask learning framework for automated and simultaneous diagnostic classification and severity assessment of COVID-19. Our experiments demonstrated that key components of the COVID-MTL framework, including unsupervised lung segmentation, Shift3D and random-weighted multitask loss, were able to improve the diagnostic workflows and outperform their counterpart methods, which finally boosted the joint learning performance when dealing with imbalanced COVID-19 tasks. Chest CT scans may contain richer information than as expected but currently not fully utilized in COVID-19 research. Including cross-continental COVID-19 CT data may further validate and improve the performance of COVID-MTL and enable it more viable for clinical practice. All our experimental data, pretrained models, and computer code can be made publicly available, which may facilitate the community for future research and relevant applications.

## Data availability and experimental reproducibility

The dataset (including 1,329 segmented chest CT scans and corresponding extracted high-throughput lung features), source code, and pretrained models will be released at a public repository upon acceptance of the paper.

## Declaration of Competing Interest

The authors declare no competing interests.

## References

[1] N. Zhu, D. Zhang, W. Wang, X. Li, B. Yang, J. Song, X. Zhao, B. Huang, W. Shi, R. Lu, P. Niu, F. Zhan, X. Ma, D. Wang, W. Xu, G. Wu, G.F. Gao, W. Tan, A Novel Coronavirus from Patients with Pneumonia in China, 2019, New England Journal of Medicine, 382 (2020) 727-733.
[2] C. Huang, Y. Wang, X. Li, L. Ren, J. Zhao, Y. Hu, L. Zhang, G. Fan, J. Xu, X. Gu, Z. Cheng, T. Yu, J. Xia, Y. Wei, W. Wu, X. Xie, W. Yin, H. Li, M. Liu, Y. Xiao, H. Gao, L. Guo, J. Xie, G. Wang, R. Jiang, Z. Gao, Q. Jin, J. Wang, B. Cao, Clinical features of patients infected with 2019 novel coronavirus in Wuhan, China, The Lancet, 395 (2020) 497-506.
[3] W.H. Organization, Considerations in the investigation of cases and clusters of COVID-19: interim guidance, 22 October 2020, World Health Organization2020.
[4] A. Tahamtan, A. Ardebili, Real-time RT-PCR in COVID-19 detection: issues affecting the results, Taylor & Francis2020.
[5] E. Surkova, V. Nikolayevskyy, F. Drobniewski, False-positive COVID-19 results: hidden problems and costs, The Lancet Respiratory Medicine.
[6] A.T. Xiao, Y.X. Tong, S. Zhang, False negative of RT-PCR and prolonged nucleic acid conversion in COVID-19: Rather than recurrence, Journal of Medical Virology, 92 (2020) 1755-1756.
[7] S. Woloshin, N. Patel, A.S. Kesselheim, False Negative Tests for SARS-CoV-2 Infection — Challenges and Implications, New England Journal of Medicine, 383 (2020) e38.
[8] S. Haseli, N. Khalili, M. Bakhshayeshkaram, M. Sanei Taheri, Y. Moharramzad, Lobar Distribution of COVID-19 Pneumonia Based on Chest Computed Tomography Findings; A Retrospective Study, Arch Acad Emerg Med, 8 (2020) e55-e55.
[9] M.-Y. Ng, E.Y. Lee, J. Yang, F. Yang, X. Li, H. Wang, M.M.-s. Lui, C.S.-Y. Lo, B. Leung, P.-L. Khong, C.K.-M. Hui, K.-y. Yuen, M.D. Kuo, Imaging Profile of the COVID-19 Infection: Radiologic Findings and Literature Review, Radiology: Cardiothoracic Imaging, 2 (2020) e200034.
[10] Q. Hu, H. Guan, Z. Sun, L. Huang, C. Chen, T. Ai, Y. Pan, L. Xia, Early CT features and temporal lung changes in COVID-19 pneumonia in Wuhan, China, European Journal of Radiology, (2020) 109017.
[11] X. Xie, Z. Zhong, W. Zhao, C. Zheng, F. Wang, J. Liu, Chest CT for Typical Coronavirus Disease 2019 (COVID-19) Pneumonia: Relationship to Negative RT-PCR Testing, Radiology, 296 (2020) E41-E45.
[12] S.A. Harmon, T.H. Sanford, S. Xu, E.B. Turkbey, H. Roth, Z. Xu, D. Yang, A. Myronenko, V. Anderson, A. Amalou, M. Blain, M. Kassin, D. Long, N. Varble, S.M. Walker, U. Bagci, A.M. Ierardi, E. Stellato, G.G. Plensich, G. Franceschelli, C. Girlando, G. Irmici, D. Labella, D. Hammoud, A. Malayeri, E. Jones, R.M. Summers, P.L. Choyke, D. Xu, M. Flores, K. Tamura, H. Obinata, H. Mori, F. Patella, M. Cariati, G. Carrafiello, P. An, B.J. Wood, B. Turkbey, Artificial intelligence for the detection of COVID-19 pneumonia on chest CT using multinational datasets, Nature Communications, 11 (2020) 4080.
[13] S. Zhou, T. Zhu, Y. Wang, L. Xia, Imaging features and evolution on CT in 100 COVID-19 pneumonia patients in Wuhan, China, Eur Radiol, 30 (2020) 5446-5454.
[14] L. Li, L. Qin, Z. Xu, Y. Yin, X. Wang, B. Kong, J. Bai, Y. Lu, Z. Fang, Q. Song, K. Cao, D. Liu, G. Wang, Q. Xu, X. Fang, S. Zhang, J. Xia, J. Xia, Using Artificial Intelligence to Detect COVID-19 and Community-acquired Pneumonia Based on Pulmonary CT: Evaluation of the Diagnostic Accuracy, Radiology, 296 (2020) E65-E71.
[15] L. Sun, Z. Mo, F. Yan, L. Xia, F. Shan, Z. Ding, B. Song, W. Gao, W. Shao, F. Shi, Adaptive feature selection guided deep forest for covid-19 classification with chest ct, IEEE

Journal of Biomedical and Health Informatics, (2020).
[16] L. Wang, Z.Q. Lin, A. Wong, COVID-Net: a tailored deep convolutional neural network design for detection of COVID-19 cases from chest X-ray images, Scientific Reports, 10 (2020) 19549.
[17] Z. Wang, Q. Liu, Q. Dou, Contrastive Cross-Site Learning With Redesigned Net for COVID-19 CT Classification, IEEE Journal of Biomedical and Health Informatics, 24 (2020) 2806-2813.
[18] M. Shorfuzzaman, M.S. Hossain, MetaCOVID: A Siamese neural network framework with contrastive loss for n-shot diagnosis of COVID-19 patients, Pattern Recognition, (2020) 107700.
[19] Z. Wang, Y. Xiao, Y. Li, J. Zhang, F. Lu, M. Hou, X. Liu, Automatically discriminating and localizing COVID-19 from community-acquired pneumonia on chest X-rays, Pattern recognition, 110 (2021) 107613-107613.
[20] Z. Tang, W. Zhao, X. Xie, Z. Zhong, F. Shi, J. Liu, D. Shen, Severity assessment of COVID-19 using CT image features and laboratory indices, Physics in Medicine & Biology, (2020).
[21] W. Ning, S. Lei, J. Yang, Y. Cao, P. Jiang, Q. Yang, J. Zhang, X. Wang, F. Chen, Z. Geng, L. Xiong, H. Zhou, Y. Guo, Y. Zeng, H. Shi, L. Wang, Y. Xue, Z. Wang, Open resource of clinical data from patients with pneumonia for the prediction of COVID-19 outcomes via deep learning, Nature Biomedical Engineering, (2020).
[22] H. Shaziya, K. Shyamala, R. Zaheer, Automatic Lung Segmentation on Thoracic CT Scans Using U-Net Convolutional Network,  2018 International Conference on Communication and Signal Processing (ICCSP)2018, pp. 0643-0647.
[23] Z. Xiao, B. Liu, L. Geng, F. Zhang, Y. Liu, Segmentation of Lung Nodules Using Improved 3D-UNet Neural Network, Symmetry, 12 (2020) 1787.
[24] H. Shi, X. Han, N. Jiang, Y. Cao, O. Alwalid, J. Gu, Y. Fan, C. Zheng, Radiological findings from 81 patients with COVID-19 pneumonia in Wuhan, China: a descriptive study, The Lancet Infectious Diseases, 20 (2020) 425-434.
[25] A. Oulefki, S. Agaian, T. Trongtirakul, A. Kassah Laouar, Automatic COVID-19 Lung Infected Region Segmentation and Measurement Using CT-Scans Images, Pattern recognition, (2020) 107747-107747.
[26] D.-P. Fan, T. Zhou, G.-P. Ji, Y. Zhou, G. Chen, H. Fu, J. Shen, L. Shao, Inf-Net: Automatic COVID-19 Lung Infection Segmentation from CT Images, IEEE Transactions on Medical Imaging, (2020).
[27] ArnavJain, K.S. Mader, DSB Lung Segmentation Algorithm (Candidate Generation and LUNA16 preprocessing: https://www.kaggle.com/arnavkj95/candidate-generation-and-luna16-preprocessing),  Data Science Bowl 2017, Kaggle2017.
[28] S. Ji, W. Xu, M. Yang, K. Yu, 3D Convolutional Neural Networks for Human Action Recognition, IEEE Transactions on Pattern Analysis and Machine Intelligence, 35 (2013) 221-231.
[29] L. Ge, H. Liang, J. Yuan, D. Thalmann, 3d convolutional neural networks for efficient and robust hand pose estimation from single depth images,  Proceedings of the IEEE Conference on Computer Vision and Pattern Recognition2017, pp. 1991-2000.
[30] H. Lu, H. Wang, Q. Zhang, S.W. Yoon, D. Won, A 3D Convolutional Neural Network for Volumetric Image Semantic Segmentation, Procedia Manufacturing, 39 (2019) 422-428.
[31] O. Köpüklü, N. Kose, A. Gunduz, G. Rigoll, Resource Efficient 3D Convolutional Neural Networks, 2019 IEEE/CVF International Conference on Computer Vision Workshop (ICCVW), (2019) 1910-1919.
[32] J. Hu, L. Shen, G. Sun, Squeeze-and-excitation networks, Proceedings of the IEEE conference on computer vision and pattern recognition2018, pp. 7132-7141.
[33] Y. Zhang, Q. Yang, A survey on multi-task learning, arXiv preprint arXiv:1707.08114, (2017).
[34] S. Ruder, An Overview of Multi-Task Learning in Deep Neural Networks, ArXiv, abs/1706.05098 (2017).
[35] R. Caruana, Multitask Learning, Machine Learning, 28 (1997) 41-75.
[36] M. Guo, A. Haque, D.-A. Huang, S. Yeung, L. Fei-Fei, Dynamic task prioritization for multitask learning,  Proceedings of the European Conference on Computer Vision (ECCV)2018, pp. 270-287.
[37] S. Liu, E. Johns, A.J. Davison, End-to-end multi-task learning with attention,  Proceedings of the IEEE Conference on Computer Vision and Pattern Recognition2019, pp. 1871-1880.
[38] A. Kendall, Y. Gal, R. Cipolla, Multi-task learning using uncertainty to weigh losses for scene geometry and semantics, Proceedings of the IEEE conference on computer vision and pattern recognition2018, pp. 7482-7491.
[39] X. Tian, Y. Li, T. Liu, X. Wang, D. Tao, Eigenfunction-Based Multitask Learning in a Reproducing Kernel Hilbert Space, IEEE Transactions on Neural Networks and Learning Systems, 30 (2019) 1818-1830.
[40] B. Bayar, M.C. Stamm, Augmented convolutional feature maps for robust CNN-based camera model identification, 2017 IEEE International Conference on Image Processing (ICIP)2017, pp. 4098-4102.
[41] Z. Chen, Y. Fu, Y. Zhang, Y. Jiang, X. Xue, L. Sigal, Multi-Level Semantic Feature Augmentation for One-Shot Learning, IEEE Transactions on Image Processing, 28 (2019) 4594-4605.
[42] H. Feng, Y. Liu, M. Lv, J. Zhong, A case report of COVID-19 with false negative RT-PCR test: necessity of chest CT, Jpn J Radiol, 38 (2020) 409-410.
[43] A.L. Mueller, M.S. McNamara, D.A. Sinclair, Why does COVID-19 disproportionately affect older people?, Aging (Albany NY), 12 (2020) 9959-9981.
[44] S. Tosif, M.R. Neeland, P. Sutton, P.V. Licciardi, S. Sarkar, K.J. Selva, L.A.H. Do, C. Donato, Z. Quan Toh, R. Higgins, C. Van de Sandt, M.M. Lemke, C.Y. Lee, S.K. Shoffner, K.L. Flanagan, K.B. Arnold, F.L. Mordant, K. Mulholland, J. Bines, K. Dohle, D.G. Pellicci, N. Curtis, S. McNab, A. Steer, R. Saffery, K. Subbarao, A.W. Chung, K. Kedzierska, D.P. Burgner, N.W. Crawford, Immune responses to SARS-CoV-2 in three children of parents with symptomatic COVID-19, Nature Communications, 11 (2020) 5703.
[45] S.P. Weisberg, T.J. Connors, Y. Zhu, M.R. Baldwin, W.-H. Lin, S. Wontakal, P.A. Szabo, S.B. Wells, P. Dogra, J. Gray, E. Idzikowski, D. Stelitano, F.T. Bovier, J. Davis-Porada, R. Matsumoto, M.M.L. Poon, M. Chait, C. Mathieu, B. Horvat, D. Decimo, K.E. Hudson, F.D. Zotti, Z.C. Bitan, F. La Carpia, S.A. Ferrara, E. Mace, J. Milner, A. Moscona, E. Hod, M. Porotto, D.L. Farber, Distinct antibody responses to SARS-CoV-2 in children and adults across the COVID-19 clinical spectrum, Nature Immunology, (2020).